\documentclass{ws-procs9x6-cpt16}
\begin{document}

\newcommand{\refeq}[1]{(\ref{#1})}
\def\etal {{\it et al.}}
\def\btpi{\boldsymbol{\tilde{\pi}}}
\def\btau{\boldsymbol{\tau}}
\def\bpi{\boldsymbol{\pi}}

\title{Testing Lorentz Symmetry using Chiral Perturbation Theory}

\author{J.P.\ Noordmans}

\address{CENTRA, Departamento de F\'isica, Universidade do Algarve\\ 
8005-139 Faro, Portugal}

\begin{abstract}
We consider the low-energy effects of a selected set of Lorentz- and CPT-violating quark and gluon operators by deriving the corresponding chiral effective lagrangian. Using this effective lagrangian, low-energy hadronic observables can be calculated. We apply this to magnetometer experiments and derive the best bounds on some of the Lorentz-violating coefficients. We point out that progress can be made by studying the nucleon-nucleon potential, and by considering storage-ring experiments for deuterons and other light nuclei.
\end{abstract}

\bodymatter

\section{Introduction}

Studies of Lorentz violation (LV) and CPT violation (CPTV) arguably are best performed in the context of the general effective-field-theory (EFT) approach called the Standard-Model Extension (SME).\cite{sme} The SME lagrangian contains all LV and CPTV operators that one can construct from the conventional Standard-Model fields, coupled to fixed-valued Lorentz tensors which parametrize the symmetry breaking and whose values presumably arise from a more fundamental theory.

One of the great merits of the SME is that it provides a concrete and general way for experiments to constrain LV and CPTV.\cite{datatables} In this respect, clock-comparison, maser, and magnetometer experiments have been particularly successful. As a result very stringent bounds have been placed on a number of effective neutron and proton coefficients.\cite{datatables} However, the connection between these effective coefficients and the quark and gluon coefficients from which they must arise is obscured by the fact that QCD is nonperturbative at the relevant (low) energies. 

In the usual Lorentz-symmetric case, the low-energy effects of QCD are addressed successfully by the EFT of QCD, called chiral perturbation theory ($\chi$PT) (for reviews see, e.g., Ref.~\refcite{chiPTreviews,chiPTreviews2}). It thus seems natural to extend this approach to the QCD sector of the SME. The conceptual basis of this extension has been established in our recent paper\cite{LVchiPTpaper} in which we treated three CPTV and LV quark and gluon operators of mass-dimension five, given by
\begin{equation}
\mathcal{L}^{\rm LV} = \sum_{q=u,d}\left[ C^q_{\mu\nu\rho}\bar{q}\gamma^\mu G^{\rho\nu} q + D^q_{\mu\nu\rho}\bar{q}\gamma^\mu\gamma^5 G^{\rho\nu}q\right] + H_{\mu\nu\rho}\,{\rm Tr}\left(G^{\mu\lambda}D^\nu\tilde{G}^{\rho}_{\;\;\;\lambda}\right)\ .
\label{lagr1gevQ}
\end{equation}
Here, the LV coefficient $H_{\mu\nu\rho}$ is fully symmetric on its Lorentz indices and $X^q_{\mu\nu\rho} = X^q_{\rho\nu\mu}$, with $X \in \{C,D\}$. These operators were first written down in Ref.~\refcite{Bol05} and are relevant below the electroweak scale $\Lambda_F \simeq 250$~GeV, but above the QCD chiral-symmetry breaking scale $\Lambda_\chi \simeq 2\pi F_\pi \sim 1$~GeV, where $F_\pi \simeq 185$~MeV is the pion decay constant. The natural expected size of the LV coefficients is $\Lambda^{-1}_{\rm LV}$, where $\Lambda_{\rm LV}$ is the scale at which the LV occurs and is commonly identified with the Planck scale $\Lambda_{\rm PL} \simeq 10^{19}$~GeV.  Because we are interested in observables for non-strange baryons, we have restricted attention to up and down quarks.

In the following we outline the approach described in Ref.~\refcite{LVchiPTpaper}.

\section{Chiral perturbation theory}

To ensure that observables that follow from an EFT correspond to the ones that arise from the underlying theory, one has to write the most general lagrangian in terms of the relevant low-energy degrees of freedom that is consistent with all the symmetries of the underlying theory.\cite{weinbergtheorm} In the resulting infinite tower of operators, operators of higher mass dimensionality are suppressed by higher powers of a large mass scale, such that for a given required accuracy the set of terms can be truncated. The suppressing mass scale determines the range of validity for the EFT.

In the case of $\chi$PT, this mass scale is given by $\Lambda_\chi$ and the relevant degrees of freedom at energies below $\Lambda_\chi$ are the pions, nucleons, and photons. Particularly important in formulating the effective lagrangian is the approximate global $SU(2)_L \times SU(2)_R \sim SO(4)$ chiral symmetry of QCD. This symmetry is explicitly broken by the small quark masses while inspection of the hadron spectrum reveals that it is also spontaneously broken down to its $SO(3)$ isospin subgroup. The pions are subsequently identified as the pseudo-Goldstone bosons. In the chiral limit (vanishing quark masses) the pions only interact through spacetime derivatives. This allows for a perturbative expansion of hadronic observables in terms of the small parameter $p/\Lambda_\chi$, where $p$ is the typical momentum of the process under consideration.

In Ref.~\refcite{LVchiPTpaper} we constructed the effective lagrangian in the $SO(4)$ formalism of $\chi$PT. The form of the effective operators is severely constrained by the transformation properties of the operators in Eq.~\eqref{lagr1gevQ} under the chiral $SO(4)$ group as well as under discrete C, P, and T transformations. The resulting set of non-redundant leading-order nucleon and pion-nucleon operators is given by
\begin{eqnarray}
\mathcal{L}^{\rm LV}_{\chi} &=& \frac{i}{m_N}\bar{N}\left(\tilde{C}^{+}_{\mu\nu\rho} + \tilde{C}^{-}_{\mu\nu\rho}\left[\tau_3 - \frac{2}{F_\pi^2 D}\left(\bpi^2\tau_3  - \pi_3 \btau\cdot\bpi\right)\right]\right)\sigma^{\nu\rho}\mathcal{D}^\mu N \notag \\
&& + \frac{1}{m_N^2}\tilde{H}_{\mu\nu\rho}\bar{N}\gamma^\mu \gamma^5 \mathcal{D}^\nu \mathcal{D}^\rho N + \frac{i}{m_N F_\pi D} \tilde{D}^-_{\mu\nu\rho}\bar{N}(\boldsymbol{\tau}\times\boldsymbol{\pi})_3\sigma^{\nu\rho}\mathcal{D}^\mu N \notag \\
&& + \mathrm{h.c.}\ ,
\label{effLagr}%
\end{eqnarray}
where $N$ is the nucleon doublet, $m_N$ is the nucleon mass, $\boldsymbol{\pi}$ is the pion triplet, $\tau_a$ are the Pauli matrices, $D = 1 + \boldsymbol{\pi}^2/F_\pi^2$, and $\mathcal{D}^\mu N$ is the chiral- and gauge-covariant nucleon derivative, whose explicit form can be found in Ref.~\refcite{LVchiPTpaper}. The LV coefficients are given by $\tilde{C}^\pm_{\mu\nu\rho} = c^\pm(C^u \pm C^d)_{\mu\nu\rho}$, $\tilde{D}^-_{\mu\nu\rho} = d^-(D^u - D^d)_{\mu\nu\rho}$, and $\tilde{H}_{\mu\nu\rho} = h H_{\mu\nu\rho}$, with $c^\pm$, $d^-$, and $h$ being low-energy constants whose size cannot be determined by symmetry considerations, although chiral symmetry gives $d^- = 2c^-$. Using naive dimensional analysis,\cite{NDA} an order-of-magnitude estimation for the low-energy constants can be obtained: $c^\pm = \mathcal{O}(\Lambda_\chi F_\pi)$ and $h = \mathcal{O}(\Lambda_\chi^2)$.

\section{Observables and limits}
The strongest bounds can be set on kinetic nucleon terms, using magnetometer experiments. These kinetic terms induce a shift in the energy levels of an atom with total angular momentum $F$ and projection number $M_F$. In Ref.~\refcite{LVchiPTpaper} we determined that the dominant part of this shift is given by
\begin{equation}
\delta E(F,M_F) = -\frac{2M_F}{F}\sum_w \sum_{N=1}^{N_w}  \left[H^{300}+(\tilde{C}^w)^{012}-(\tilde{C}^w)^{021}\right]\left\langle [\sigma_3]_{w,N}\right\rangle\ ,
\end{equation}
where $w$ labels the particle species (neutron or proton) and $N$ runs over all particles of that species that are present in the atom. The unknown matrix elements $\left\langle [\sigma_3]_{w,N}\right\rangle$ are defined in the `stretched' state $|F,F\rangle$ and one has to adopt some nuclear-structure model to determine their value. Using the results of a $^3$He/$^{129}$Xe comagnetometer experiment,\cite{comagnetometer} we were able to put the current best bounds on components of $C$ and $H$, on the order of $10^{-33}$~GeV$^{-1}$. Subleading effects of $C$ and $H$ on atomic energy levels, as well as their effect on spin-precession, can be found in Ref.~\refcite{LVchiPTpaper}. 

The LV $D$ coefficient does not show up in these calculations, because it has no kinetic nucleon terms in Eq.~\eqref{effLagr}. Such terms are forbidden by the symmetries. In Ref.~\refcite{LVchiPTpaper}, we calculated the contribution of $D$ to the electromagnetic form factor of the nucleon, through pion loops. Experimental effects of this are suppressed with a factor of $10^{-15}$ with respect to effects of $C$ and $H$, due to small electromagnetic fields. This leads to a best bound on some components of $D$ in the order of $10^{-18}$~GeV$^{-1}$, which does not probe the expected size of $1/\Lambda_{\rm PL}$. However, despite the lack of kinetic nucleon terms, the $D$ coefficient does contribute to one-pion exchange between nucleons.\cite{LVchiPTpaper} In future work it might therefore be possible to improve the bounds on $D$ by considering the LV nucleon-nucleon (NN) potential. Studies of the NN potential and its applications to deuterons and other light nuclei have been performed successfully in the context of chiral EFT with discrete-symmetry violations other than CPTV (see Ref.~\refcite{reviewNN} for a review). An analogous treatment in the present case is expected to lead to new bounds from atomic experiments or storage-ring experiments for the deuteron or other light nuclei.

\section*{Acknowledgments}
The author acknowledges financial support from the Portuguese Foundation for Science and Technology (FCT) under grant SFRH/BPD/101403/2014 and program POPH/FSE.

\end{document}